\documentclass[twocolumn,showpacs]{revtex4}

\usepackage{graphicx}
\usepackage{dcolumn}
\usepackage{bm}
\usepackage{amsmath}
\usepackage{amssymb}

\begin{document}

\title{Analogue Gravity and ultrashort laser pulse filamentation.}

\author{D. Faccio$^{1}$, S. Cacciatori$^{2}$, V. Gorini$^{2}$, V.G. Sala$^{1}$, A. Averchi$^{1}$, A. Lotti$^{1}$, M. Kolesik$^3$, J.V. Moloney$^3$}

\affiliation{$^1$CNISM and Department of Physics and Mathematics, Universit\`a dell'Insubria, Via Valleggio 11, I-22100 Como, Italy\\
$^2$Department of Physics and Mathematics, Universit\`a dell'Insubria, Via Valleggio 11, I-22100 Como, Italy\\
$^3$Arizona Center for Mathematical Sciences and Optical Sciences Center,
University of Arizona, Tucson, 85721 AZ, U.S.A}
\date{\today}

\begin{abstract}
Ultrashort laser pulse filaments in dispersive nonlinear Kerr media induce a moving refractive index perturbation which modifies the space-time
geometry as seen by co-propagating light rays. We study the analogue geometry induced by the filament and show that one of the most evident
features of filamentation, namely conical emission, may be precisely reconstructed from the geodesics. We highlight the existence of favorable
conditions for the study of analogue black hole kinematics and Hawking type radiation.
\end{abstract}

\maketitle

The persistent lack of a satisfactory quantum model of gravity has led to the development of a semiclassical approach
in which the dynamics of a quantum field is analyzed in the curved spacetime metric determined by the classical gravitational field \cite{birrell}.
One of the most intriguing predictions of the semiclassical model is the thermalization of the quantum field
in the presence of the event horizon of a black hole \cite{hawking}. However, it turns out that for a typical stellar mass black hole
the temperature of this expected radiation is so low ($\sim10$ nK) and the intensity of the latter far away from the actual source is so weak as to be extremely
unlikely to ever be detected. On the other hand, Unruh proposed
the use of gravity analogue systems that reproduce the kinematics of gravitational systems as a means to gain insight and experimental verification
of gravitational theories in earth-based laboratories \cite{unruh}. A number of analogue systems have been proposed, relying mainly on acoustic
perturbations in low-temperature condensates and on light rays in a moving dielectric \cite{novello_book,barcelo_review}. These mostly suffer
from the same limitation of astrophysical black holes, i.e. extremely low Hawking temperatures, and therefore continue to pose a strong experimental
challenge. More recently Philbin et al. proposed a dielectric medium based analogue in which a fiber soliton generates, through the nonlinear Kerr effect, a
refractive index perturbation (RIP) which modifies the space-time geometry as seen by co-propagating light rays \cite{philbin}. The Hawking
temperature in the laboratory reference frame is found to be given by $kT=\hbar\alpha/2\pi c$ where the
surface gravity $\alpha=-(c/\delta n)\partial \delta n/\partial \tau$ is determined by the variation of the RIP along the retarded time coordinate
$\tau$.\\
Ultrashort laser pulse filamentation in bulk dispersive media with Kerr nonlinearity is characterized by the transformation of an input bell-shaped
laser pulse into a tightly focused, high intensity peak that may propagate sub-diffractively over distances much larger than the Rayleigh length
\cite{couairon_review}.
One of the most spectacular manifestations of filamentation is conical emission, i.e. the generation of new wavelengths propagating along a cone
with the cone angle typically increasing with increasing wavelength shift from the pump pulse. A further important feature of filamentation
that is relevant for what follows is the typical pulse splitting process that leads to the formation of two daughter pulses, one traveling
slightly slower, the other slightly faster than the input Gaussian pulse.\\
\indent In this letter we present a dielectric analogue based on the RIP induced by ultrashort laser pulse filamentation. We analyze in detail the
light-ray geodesics predicted for such a system and show that these lead to a conical emission that is described by the same relation
predicted by other models. This is verified experimentally in the present work both for a narrow bandwidth probe pulse and for a spectrally broad pulse. In the latter
case a blue-shifted and a red-shifted spectral peak appear and this is interpreted as evidence of the complex Doppler effect \cite{frank,sorokin}.
Finally we propose possible operating conditions for the observation of Hawking radiation.\\
\indent We start by analyzing how the filament modifies the effective space-time geometry as seen by light rays. The intense core of the filament excites,
through the nonlinear Kerr effect, a refractive index profile $n(x,y,z,t)=n_0+\delta n(x,y,z,t)$ where $\delta n(x,y,z,t)=n_2I(x,y,z,t)$, $n_2$
is the nonlinear Kerr coefficient and $I(x,y,z,t)$ is the intensity profile of the filament light pulse. We note that here $z$ is the pulse
longitudinal coordinate directed along the propagation direction and $t$ is the time in the laboratory reference frame. The analogue space-time metric in the laboratory
reference frame is
\begin{equation} \label{metric}
ds^2=\cfrac{c^2}{n(x,y,z,t)^2} dt^2- dx^2 -dy^2 -dz^2=g_{\mu\nu} dx^\mu dx^\nu.
\end{equation}
Note that this is not the metric describing the light propagation in a moving medium, see \cite{Leonhardt:1999fe}.
This is because, in our case, even though the index profile is
moving, the dipoles constituting the medium are at rest in the laboratory.
The null geodesics in the laboratory reference frame are the solutions of the geodesic equation
$\ddot x^\mu +\Gamma^{\mu}_{\nu\sigma} (x(\lambda)) \dot x^\nu \dot x^\sigma=0$ with $g_{\mu\nu}(x(\lambda)) \dot x^\mu (\lambda) \dot x^\mu (\lambda)=0$
and $\Gamma^{\mu}_{\nu\rho}= (1/2) g^{\mu\sigma} [\partial_\nu g_{\sigma\rho}+\partial_\rho g_{\sigma\nu}-\partial_\sigma g_{\nu\rho}]$.
\begin{figure}[!t]
\includegraphics[angle=0,width=8cm]{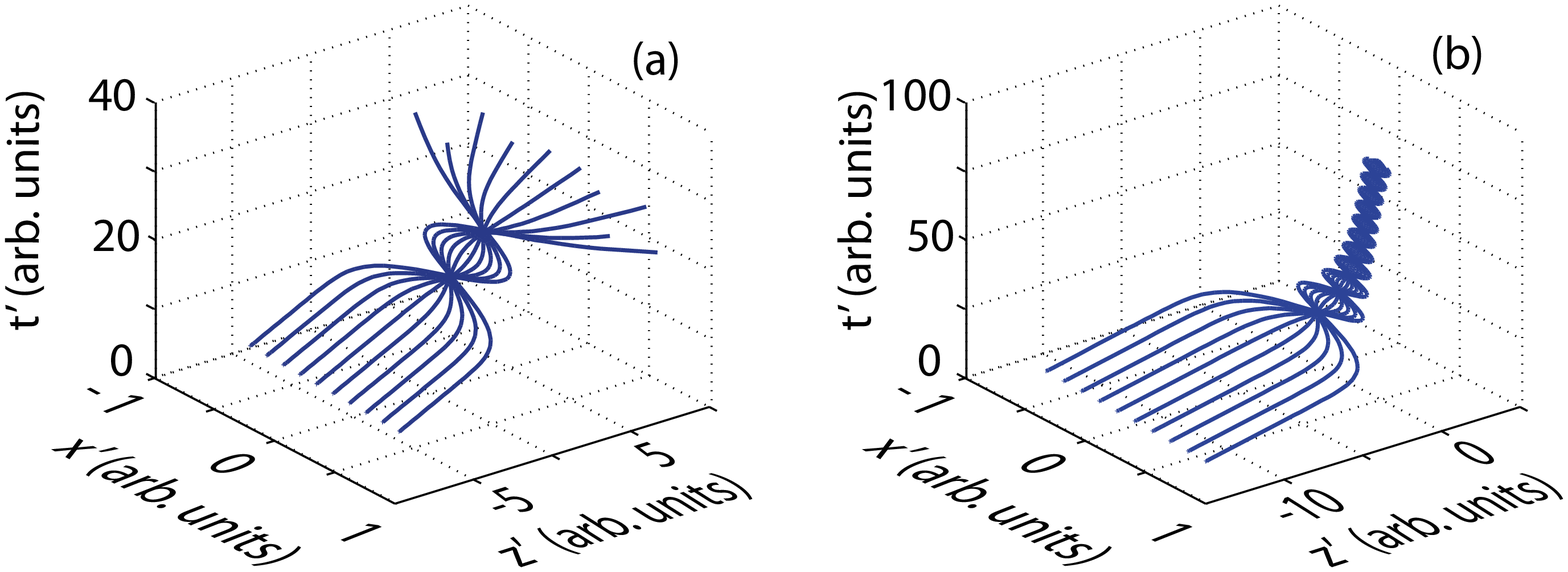}
	\caption{ \label{fig:fig1}(a) Space-time light ray geodesics in the reference frame of the moving RIP. (a) corresponds to a subluminal RIP: the light rays catch up with the RIP and the geodesics are deformed from the initial ``straight'' propagation. After two refocusing cycles (the details depend on the initial conditions) the light rays finally escape from the RIP with modified angles and frequencies. (b) corresponds to an RIP with $v=c/(1+\delta n_0)$: the light rays are now trapped by the RIP event horizon and cannot escape. The increasing steepness of the curves as they approach z=0 tells us that light is grinding to a halt inside the RIP.}
\end{figure}
Figure~\ref{fig:fig1}(a) shows the calculated geodesics in the RIP co-moving reference frame, considering a refractive index perturbation profile in the form of a Gaussian,
$\delta n=\delta n_0\exp[{-(z-vt)^2-x^2-y^2}]$. The RIP bends the
geodesics toward its center and introduces also a time deformation. This leads to a simultaneous
modification of the light ray propagation angle $\theta$ and frequency, $\omega$. These effects are best studied in $(\theta,\omega)$ space.
We start from the incoming light ray wavevector which may be written in the RIP co-moving reference frame as
$k_{\rm in}=(k^{0\prime}_{\rm in}, 0,0,k'_{\rm in})$. The spacetime metric experienced by the observer co-moving with the filament is
\begin{eqnarray} \label{metricprime}
ds^2&=&c^2\left( 1+\left[\frac {1}{ n^2} -1\right]\gamma^2 \right)dt'^2 +2 \gamma^2 v \left[\frac 1{ n^2} -1\right]dt'\ dz'\ \nonumber \\
&-&\left( 1-\left[\frac 1{ n^2} -1\right]\gamma^2 \frac {v^2}{c^2} \right)dz'^2-dx'^2-dy'^2 \nonumber \\
&=& g'_{\mu\nu} dx^{\prime\mu} dx^{\prime\nu},
\end{eqnarray}
where $\gamma=1/\sqrt{1-(v/c)^2}$.\\
Note that in this frame the metric is stationary, being independent from the coordinate $t'$. Then, time translations are symmetries and
determine a conserved quantity: $E=k^{\prime \mu} \xi^{\prime\nu} g'_{\mu\nu}$, where $\xi'=(1,0,0,0)$ is the associated Killing vector.
Thus $E=k^{\prime \mu} g'_{\mu0}=k'_0$ and $E_{\rm out}=E_{\rm in}$ is equivalent to
\begin{eqnarray}
k_{0 {\rm out}}^{\prime}=k_{0 {\rm in}}^{\prime}. \label{tre}
\end{eqnarray}
In other words in the comoving frame the frequency is constant along a geodesic, as expected ($\omega'=ck'_0$).
Turning back to the laboratory frame, where $k_{\rm in}=(k_{0\rm in}, 0,0, -k_{\rm in})$ and
$k_{\rm out}=(k_{0\rm out}, -k^x_{\rm out}, -k^y_{\rm out}, -k^z_{\rm out})$,
\begin{figure}[t]
\includegraphics[angle=0,width=8cm]{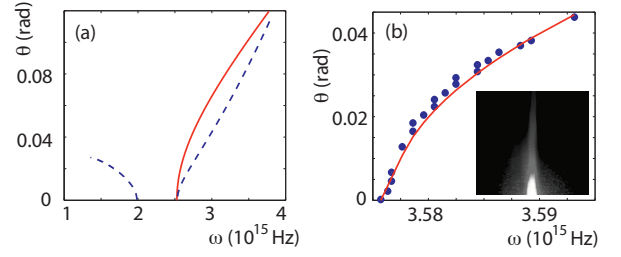}
	\caption{ \label{fig:fig2} $(\theta,\omega)$ spectrum of the scattered light rays calculated from Eq.~(\ref{formula}).
(a) solid line - neglecting material dispersion (equivalent to the standard Doppler effect), dashed line - including material dispersion
(equivalent to the complex Doppler effect). (b) Measured spectral deformation of  a probe pulse in the presence of a filament.
Circles - experimental maximal intensity points taken from the measurement shown in the inset. Solid line - fit obtained using
Eq.~(\ref{formula}) neglecting material dispersion.}
\end{figure}
equation (\ref{tre}) becomes
\begin{eqnarray}\label{formula}
k^z_{\rm out} =k_{\rm in}+\cfrac{\omega_{\rm out}-\omega_{\rm in}}{v}.
\end{eqnarray}
Finally, the $(\theta,\omega)$ distribution of the outgoing light rays as measured in the laboratory reference frame is given by inserting
Eq.~(\ref{formula}) in
$\theta\simeq k_\perp/k(\omega)=\sqrt{1-[k_{z\rm out}/k(\omega)]^2}$ where $k(\omega)$ is the material dispersion relation.
We underline that this last equation, along with Eq.~(\ref{formula}), is identical to that derived within the framework of the Effective-Three-Wave-Mixing (ETWM)
and X-wave models that have been successfully used to describe the main spectral features and dynamics of ultrashort laser pulser filamentation \cite{miro_etwm,kolesikprl,conti,faccioprl,faccioOE,aleOL}. The present derivation of these relations  on the one hand lends further support and understanding 
to these interpretations of dynamics in nonlinear optics. On the other hand, it justifies our analogue 
model as a candidate to study the analogue gravity.  It is also important to note that Eq.~(\ref{formula}) has been derived under the approximation of negligible material
dispersion {\emph{within}} the RIP. We have verified that this approximation is valid as long as the difference between the RIP velocity, $v$ and the material group velocity, $v_g=dk/d\omega$, is small. A more detailed discussion regarding this point will be given in a future publication but, as verified below, filamentation is indeed characterized by $|(v-v_g)/v_g|\ll1$, thus justifying our derivation.  \\
\indent We underline that one of the cardinal points in deriving Eq.~(\ref{formula}) is the Doppler shift that allows to transform back and forth between the two reference systems. On the basis of this, two different regimes may be highlighted depending on how one accounts for the material dispersion in the medium surrounding the RIP. If we neglect the material dispersion,
i.e. $k(\omega)=\omega n(\omega_{\rm in})/c$ we obtain the solid curve in Fig.~\ref{fig:fig2}(a). This is equivalent to the standard
Doppler effect in which observers at an angle with respect to the direction of motion of a scatterer experience a sweep in the scattered frequency. However,
along the line of motion, no frequency shift in the {\emph{forward}} direction is observed. Conversely, we may fully account for material dispersion, i.e.
$k(\omega)=\omega n(\omega)/c$. In this case we obtain the dashed curve in Fig.~\ref{fig:fig2}(a). 
This is equivalent to the so-called complex
Doppler effect \cite{frank} in which the emitter is seen to emit two (or more) frequencies along the direction of motion. Indeed, in the presence
of dispersion the Doppler formula $\omega'=\gamma\omega[1-vk(\omega)]$, which lies at the basis of the coordinate transformations used in deriving
Eq.~(\ref{formula}), may have two or more solutions depending on the complexity of the $k(\omega)$ dispersion  relation \cite{frank,sorokin}.\\
\indent We experimentally verified these findings. Experiments were carried out using a regeneratively amplified Nd:Glass laser delivering pulses
at 10 Hz repetition rate with 1 ps pulse duration and 1055 nm central wavelength (Twinkle, Light Conversion Ltd., Lithuania). The nonlinear Kerr
medium was a 2 cm long fused silica sample. A filament was generated by focusing down to a $100$ $\mu$m input diameter a pulse with an input
energy of 20 $\mu$J. A second probe pulse is obtained by loosely focusing (1 mm diameter) 100 nJ at the second harmonic wavelength, thus allowing to easily distinguish the pump and probe pulses.. This allows one to
easily separate the filament and probe pulses at the sample output.  In Fig.~\ref{fig:fig2}(b) the solid circles show points of maximal intensity
taken from the measured output probe spectrum (shown in the inset). Due to the limited spectral bandwidth of the probe pulse we may effectively
neglect material dispersion and thus obtain a good fit for the data with Eq.~(\ref{formula}), using $k(\omega)=\omega n(\omega_{\rm in})/c$.\\
\begin{figure}[t]
\includegraphics[angle=0,width=8cm]{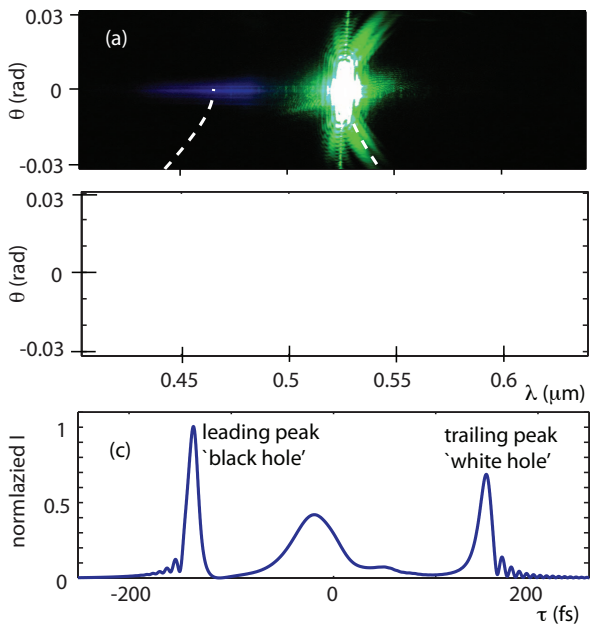}
	\caption{ \label{fig:fig3} $(\theta,\lambda)$ spectra of filament pulses in a 3 cm fused silica sample. (a) input energy =1.5 $\mu$J,
(b) input energy =1.8 $\mu$J. The dashed lines correspond to the spectral supports predicted by Eq.~(\ref{formula}), including material
dispersion and with (a)  $v=0.996c/n_g$ and (b)  $v=1.0032c/n_g$, $v=0.997c/n_g$. (c) numerically simulated temporal profile of the filament.}
\end{figure}
In order to test the validity of Eq.~(\ref{formula}) in the presence of material dispersion we performed a second set of experiments in which the
pulse probing the RIP is the same pump pulse generating the filament. It is well-known that filamentation will lead to pulse splitting and to
significant pulse shortening \cite{couairon_review}. Starting from an input 200 fs pump pulse at the second harmonic of our laser, i.e. at 527 nm, the filament dynamics 
lead to pulse  durations  as short as 20-30 fs  with correspondingly broad spectral bandwidths. The input pulse was focused to a 100 $\mu$m diameter at the input of a 3 cm long
fused silica sample. By carefully controlling the input energy it was possible  to shift the onset of filamentation toward the end of the sample so
that only a very short filament was observed before the pulse exited into air, thus quenching all nonlinear effects and subsequent interactions.
In this way we avoid the relatively intense on-axis frequency continuum generated by long-scale filaments that would completely cover the spectral
features that we are aiming to unveil. Figure~\ref{fig:fig3}(a) shows the measured spectrum with an input energy of $1.5$ $\mu$J. The input
Gaussian pump pulse spectrum has been reshaped by the RIP and exhibits marked red-shifted conical emission tails and a new blue-shifted component,
mainly concentrated around $\theta=0$. The dashed lines show the spectral support predicted from Eq.~(\ref{formula}) with $v=0.996c/n_g$ where
$n_g=c/v_g$ is the material group index at 527 nm. The observed spectral features are well reproduced by the curve and we may thus interpret the spectrum
as a manifestation of the complex Doppler effect by which the input 527 nm pulse is scattered by an RIP that is traveling slightly slower than the
input Gaussian pulse. We note that a similar blue-shifted peak has
been previously reported \cite{blue_peak} and was shown to be related to a steep trailing front within the filament. This is also in keeping with
the results and interpretation of Philbin et al., who assimilates a trailing pulse edge to a white hole \cite{philbin}. Such a white hole will
indeed blue-shift any incoming light rays, as also observed in our measurement. \\
Figure~\ref{fig:fig3}(b) shows the measured spectrum with a slightly higher input energy of $1.8$ $\mu$J. Now, in addition to the blue-shifted
peak, we also observe a red-shifted peak. As before, the dashed lines show the spectral support predicted from Eq.~(\ref{formula}), with
$v=0.997c/n_g$ for the blue-shifted spectrum and with $v=1.0032c/n_g$ for the red-shifted spectrum. The fact that the blue peak may appear
independently of the red peak excludes four-wave-mixing as the generating process. We thus interpret the on-axis red peak as the result of the
complex Doppler effect for light rays scattered from a leading front in the filament.  This leading front is just a time-reversed realization of
the trailing front analyzed in Fig.~\ref{fig:fig3}(b) and it may thus be assimilated to a black hole \cite{philbin}. In Fig.~\ref{fig:fig3}(c)
we show the numerically calculated intensity profile of the pulse along the propagation axis at the sample output. A description of the numerical code
is given in \cite{miro_uppe,miro_uppe2,miro_etwm} and a detailed comparison with experiments is given in \cite{blue_peak}. As can be seen, the pulse is indeed characterized by a sharp trailing edge, that gives rise to
a blue-shifted spectral peak, and by an equivalently sharp leading edge that gives rise to red-shifted spectral peak. For different input energies
the filament dynamics may change significantly and only a trailing pulse (and blue shifted spectral peak) may be observed  \cite{blue_peak},
thus explaining the difference between Figs.~\ref{fig:fig3}(a) and (b).
We note that for a pulse propagating in nonlinear media with a positive Kerr coefficient, the formation of a shock front on the trailing edge is
to be expected and has been frequently reported \cite{shocks,alfano_book}. However the formation of a leading shock front is not as obvious and
indeed is a peculiar feature of filamentation \cite{shockedX}. This aspect makes filamentation particularly suited to the study of 
what concerns the present work.\\
\indent We underline once more that in order to actually observe Hawking radiation the RIP should travel superluminally with respect to
the phase velocity of the incoming light rays. Figure~\ref{fig:fig1}(b) shows the geodesics in the co-moving RIP reference frame in the limiting case for the horizon event formation, i.e. for an RIP velocity equal to the light phase velocity. The geodesics are now strongly deformed and, as light gradually grinds to a halt,  cannot escape from the RIP.
From the experimental viewpoint, Fig.~\ref{fig:fig3}(c) shows how the filament pulse (and thus also the
corresponding RIP) consists of two main peaks, with the trailing peak being slower and the leading peak faster than the input pulse. It has
been shown that by correctly choosing the input pulse conditions in function of the specific Kerr medium, it is possible to create a leading
pulse that is superluminal even with respect to the same input carrier frequency \cite{cep}. Such an RIP will therefore appear as a black hole for the pump pulse carrier frequency.\\
\indent In conclusion we have described the space-time geometry of light rays in the presence of a filament. The equations predict a specific
shape for the aberrated light rays that has been confirmed experimentally. The results indicate the presence of both white and black hole horizons.
Using the same formulas of \cite{philbin}, a 5 fs pulse at 800 nm (2 optical cycles) will lead to a Hawking temperature corresponding to a
peak wavelength around 1.2 $\mu$m for which high sensitivity detectors are readily available. One may envision other possibilities for achieving
superluminal pulses in Kerr media such as the use of Bessel-like pulses, although filaments maintain a certain attractiveness due to the
spontaneous pulse shortening and the possibility to observe more exotic phenomena such as black hole lasing \cite{blackhole_laser1,blackhole_laser2}.\\
\indent The authors acknowledge measurements performed in the Ultrafast Nonlinear Optics Laboratory, Universit\`a dell'Insubria, Como, Italy, led by Prof. Paolo Di Trapani. JVM acknowledges support from US Air Force Office of Scientific Research (AFOSR) under
contract FA9550-07-1-0010.

\end{document}